\documentclass[aps,pra,twocolumn,floatfix]{revtex4-1}

\usepackage{amssymb}
\usepackage{amsmath}
\usepackage{graphicx}
\usepackage{dcolumn}
\usepackage{bm}
\usepackage{float}
\usepackage{color}
\usepackage{hyperref}
\hypersetup{colorlinks=true, citecolor=blue, urlcolor=blue, linkcolor=blue}
\usepackage{geometry}
\newcommand{\bra}[1]{\langle #1|}
\newcommand{\ket}[1]{|#1\rangle}
\newcommand{\braket}[2]{\langle #1|#2\rangle}

\DeclareMathOperator{\sech}{sech}

\begin{document}
\title[Populations in Bragg Regime]{ Nonadiabatic contributions to Bragg-regime dynamics\\ in atomic Kapitza-Dirac scattering}
\author{Dylan Manna}
\affiliation{University of Michigan, Ann Arbor, Michigan 48109, USA}

\begin{abstract}
Atomic Kapitza-Dirac Bragg regime scattering is a multiphoton process in which a neutral atom undergoes a change of momentum through an interaction with a coherent light source. When the Bragg conditions are met, the outgoing atom beams are spatially quantized. Counterpropagating lasers act as pump and probe scattering from far-off-resonant excited intermediate electronic states, leaving the atoms in the electronic ground state with quantized transverse momentum. Each nontrivial scattering event imparts transverse velocity and therefore kinetic energy to the deflected atoms through recoil. In the Bragg regime, the loss of energy in the light fields is equal to the gain of kinetic energy in the atom. Energy nonconserving intermediate states, which are described by nonadiabatic contributions, are not accounted for by first-order off-resonant states. By comparing the solutions of increasing orders of off-resonant intermediate states, it becomes clear that calculating the correct Pendell\"{o}sung frequencies and phases requires including an ever increasing number of off-resonant states in proportion to the square root of the field strength.\\
\\
\textbf{DOI:} \href{https://doi.org/10.1103/PhysRevA.101.063621}{10.1103/PhysRevA.101.063621}

\end{abstract}

\maketitle

\section{\label{introduction}Introduction}

Two decades after J.J. Thompson introduced Bragg scattering at a meeting of the Cambridge Physical Society \cite{bragg1912} in 1912, Kapitza and Dirac first introduced what has become known as the Kapitza-Dirac effect in 1933 \cite{kde}. Bragg \cite{bragg1912} had successfully demonstrated the coherent scattering of x rays from crystals, whereas Kapitza and Dirac, seeking analogs between matter and light, proposed the theory of coherent electron diffraction from a strong light source. The reciprocity of electrons and photons in quantum electrodynamics suggests that one could observe Bragg diffraction of matter from a periodic light source analogous to x-ray diffraction from crystalline solids \cite{batterman}. In the energy-conserving Bragg regime, the outcome of this effect is the spatially quantized diffraction of the electron beam into undeflected and deflected trajectories where the momentum and energy of the electron and light quanta would be conserved. 

Given the small coupling of light with electrons, the realization of the Kapitza-Dirac effect (KDE) required the advent of lasers. By 1965, experiments scattering electrons from lasers were proposed and successfully performed \cite{eberly,compton,mciver,batelaan2,batelaan4,batelaan3,carsten}. With the many advances in atomic optics made in the 1980s, the definition of the Kapitza-Dirac effect was expanded to include neutral atoms diffracting from periodic gratings created by counterpropagating lasers \cite{pritchard,PhysRevA.36.2495,PhysRevLett.61.1182,ketterlespinor,pritchardold,24photons,Gadway:09}. The atom-optics Bragg scattering theory was put forth in 1980 \cite{PhysRevA.23.1290} and 1985 \cite{Pritchard:85} and observed experimentally in 1987 \cite{PhysRevLett.60.515}. In contrast to the simpler photon-electron interaction, atomic Kapitza-Dirac scattering is complicated and enriched by the electronic structure of the atom and the dressed field  \cite{deflection,dynamicdiff,sancho}. Kozuma \textit{et al.} \cite{phillips} utilize the atomic KDE in order to study the behavior of Bose-Einstein condensates where spatial quantization is readily observed. Here and in other experiments which involve the atomic KDE, one may use arbitrary pulse shapes as the experiment is designed in a pump-probe arrangement where the condensate is still and the lasers are pulsed, yielding recoil momentum. Gupta \textit{et al.} \cite{gupta} examine the Bose-Einstein condensate laser which relies on the KDE as the mechanism of coherently imparting momentum.

In the reporting of atom-optic Kapitza-Dirac Bragg scattering by Pritchard \textit{et al.} \cite{PhysRevLett.60.515}, Fig. 4 contains a plot of population vs power. Although  Pendell\"{o}sung is observed, the phase and frequency deviate appreciably from the theoretical curve. Lee \textit{et al.} in 1995 \cite{giltner} refer to the discrepancies in the pioneering work of Pritchard \textit{et al.}, as well subsequent experimental work, between the predicted and measured Pendell\"{o}sung as a function of field strength in first-order Bragg scattering. Precision experiments involving spatial quantization will require better understanding of the approximation techniques and their limitations \cite{zhan,rose,quasicrystal,sengstock}.

\section{\label{doppler}Doppler-shifted Laue Geometry}

\begin{figure}[H]
\includegraphics[width=8cm]{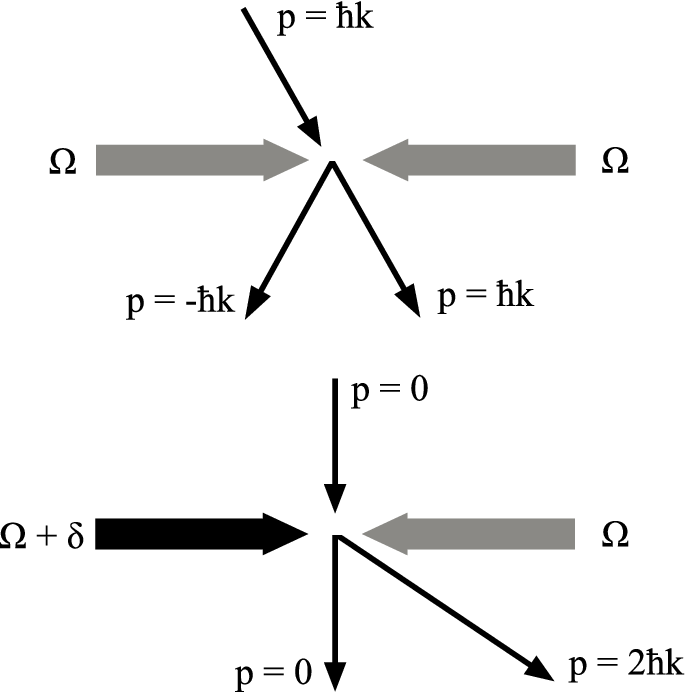}
\caption{Top: The symmetric Laue geometry. Bottom: The
Doppler-shifted Laue geometry.}
\label{fig:geometry}
\end{figure}

In the symmetric Laue geometry, the atom enters the light field at the Bragg angle and exits undeflected or at the diffracted Bragg angle. In the Doppler-shifted geometry, however, the incoming atomic beam is cooled transverse to its classical trajectory such that the angle of incidence is perpendicular to the light field as is shown in Fig. \ref{fig:geometry}. This transformed geometry requires that the counterpropagating lasers are relatively detuned to mimic the Doppler-frequency-shifted photons that the atom would see in symmetric Bragg scattering. This detuning is analogous to a system of x rays incident on moving crystals \cite{xray,buras} and has a similar advantage as the incident beam angle need not be recalibrated for varying Bragg conditions. The experimentalist sets the incoming beam angle normal to the laser propagation direction while adjusting the detector's angle and the relative laser frequencies. As field strength is varied, relative Pendell\"{o}sung in the populations of the outgoing beams is observed analogous to Pendell\"{o}sung in Bragg scattering of x rays in crystals in a Laue geometry where the field strength of the light grating is the corollary to the thickness of the crystalline solid. \cite{neutronpend}.

\begin{figure}[H]
\includegraphics[height=5cm]{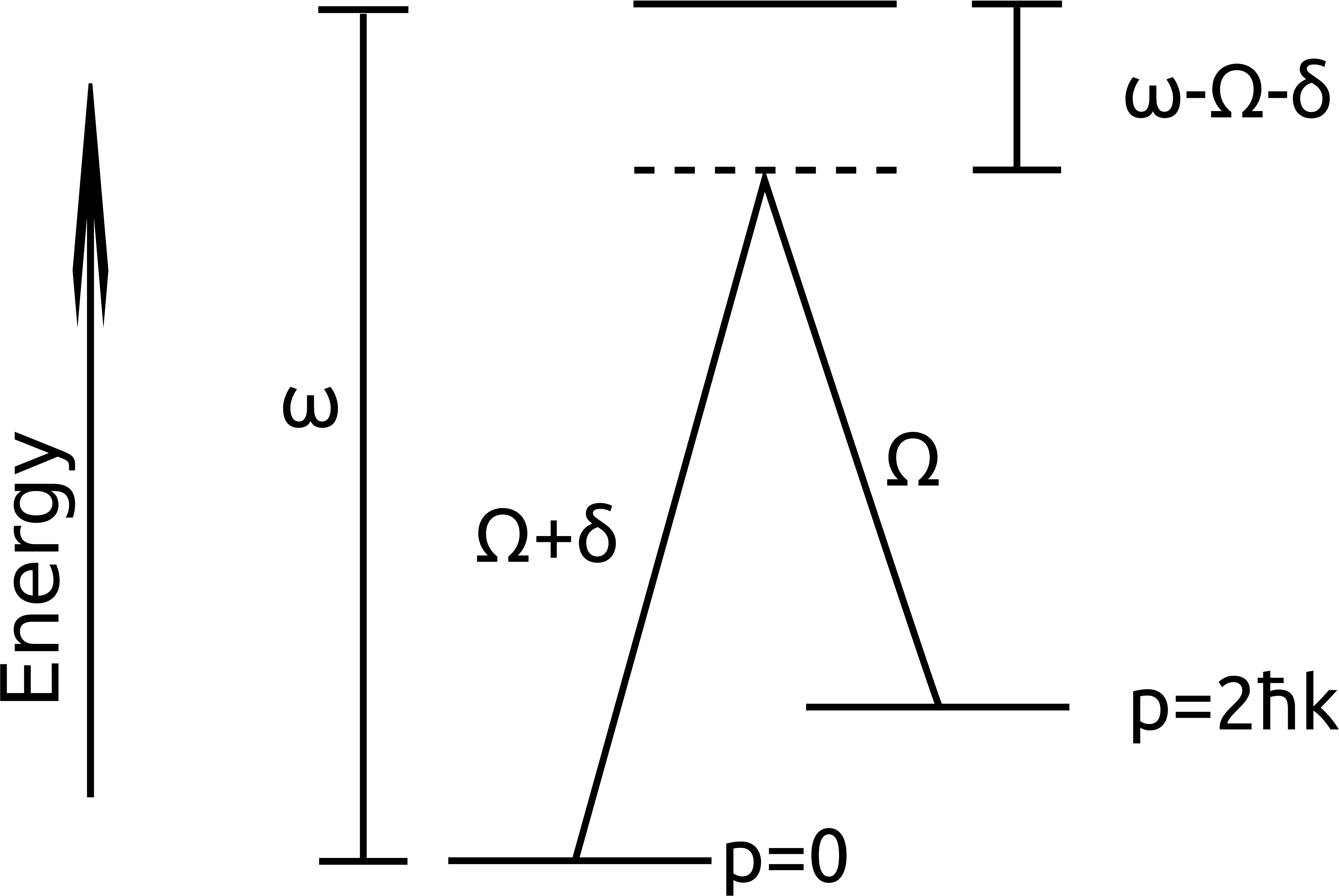}
\caption{Energy diagram showing off-resonant Raman scattering. Note that $\omega-\Omega \gg \delta$.}
\label{fig:levels}
\end{figure}

In the Doppler-shifted Laue geometry, one sets the initial transverse speed to $0$, and the field frequencies the atom sees are $\Omega+\delta$ and $\Omega$, respectively. The subsequent virtual absorption and emission processes leading to the population of scattering states with $2\hbar k$ are depicted in Fig.~\ref{fig:levels} where the Raman resonance condition is insured by matching the two-photon process detuning $\delta$ to the energy difference between the bare and scattered states. The two states differ in energy by four times the recoil frequency defined as $\omega_\text{rec}=\hbar k^2/(2m)$, where the transverse momentum is $\hbar\Omega/c+\hbar(\Omega+\delta)/c \approx 2\hbar k$. This satisfies the Bragg condition for resonant scattering into a transverse momentum state as
\begin{equation}
\delta=4\omega_\text{rec}=\frac{2\hbar\Omega^2}{mc^2}
\end{equation}

\noindent as the exchange leaves the field with an energy difference of $\hbar(\Omega+\delta)-\hbar\Omega=\hbar\delta$, equaling exactly the kinetic energy gained by the diffracted atom. In the frame of the incoming atom, the interaction is that of an atom experiencing two counterpropagating pulses with an envelope defined by the laser profile. 

\begin{figure}
\includegraphics[width=8cm]{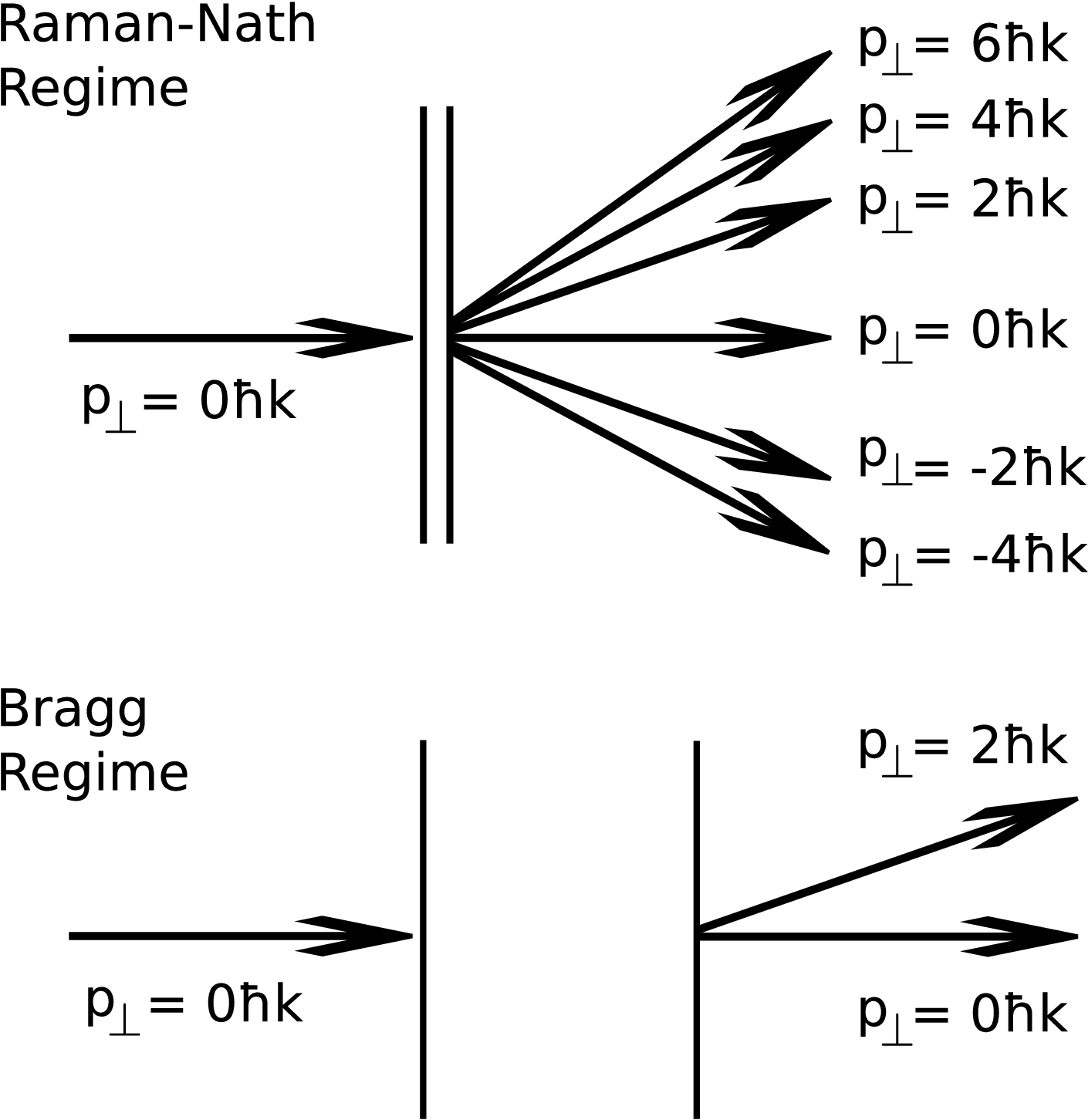}
\caption{The Raman-Nath and Bragg regimes}
\label{fig:ramanath}
\end{figure}

To obviate the effect of spontaneous emission or any other losses affecting the electronic excited state, the laser frequencies of the fields must be far-off-resonant with the internal electronic level energy $\omega$ inferring that $\omega - \Omega$ must be large enough to guarantee that no electronic atomic states are populated during the scattering event. We set the relevant single-photon detuning $\Delta = \omega - \Omega$ and we assume it in the following to be much larger than other involved rates, for example $\Delta \gg \delta$. In the frame of the incoming particle, the interaction is that of an atom experiencing two counterpropagating pulses with an envelope defined by the laser profile. We define an interaction time $\tau$ defined as the laser beam width divided by the longitudinal speed of the incoming atom beam. Generally one distinguishes between short and long interaction times based on the ratio between the two-photon detuning $\delta$ and the inverse of $\tau$. Roughly speaking, for short interaction times where the effective pulse bandwidth $\tau^{-1}>\delta$, many orders of energy nonconserving terms contribute during the interaction; that is, nonadiabatic contributions should be considered as leading to many scattering states (represented in Fig.~\ref{fig:ramanath} as the Raman-Nath regime). However, we consider in the following scattering output states strictly in the Bragg regime, i.e., energy conserving. Therefore we require that the two-photon effective pulse bandwidth be small such that one can easily resolve the Bragg scattering state, i.e., $\tau^{-1}\ll\delta$. Consequently, after the interaction the atom will be in one of two outgoing states: undeflected or deflected with an angle,
\begin{equation}
\theta=\arctan{(\frac{2\hbar k}{p_0})},
\end{equation}

\noindent where $\vec{p_0}$ is the initial momentum and $p_{0_\perp}=0$. Short time interactions in the Raman-Nath regime which allow for energy nonconserving output states, shown for comparison in Fig. \ref{fig:ramanath}, are not examined here. It is important to recall that despite being in the Bragg regime, intermittent energy nonconserving processes are permitted and contribute to a reduction of scattering into the deflected $2\hbar k$ Bragg deflected beam. We analytically and numerically discuss the contributions of such two-photon off-resonant contributions to the deflected beam amplitude and show that even under standard conditions such contributions are non-negligible.

\section{\label{raman}Virtual Two-Photon Raman Scattering}

In atomic Bragg scattering with symmetric Laue geometry, the frequency of both lasers is identical. The incoming atom has a transverse momentum that is reversed after virtual excitation and stimulated emission provides a recoil equal to twice the incoming transverse momentum. There is no change in the kinetic energy of the atom, and the counterpropagating fields, albeit respectively losing and gaining a photon, experience no net change in energy. The exchange is achieved through two-photon-stimulated Rayleigh scattering as the atom merely changes direction. In this symmetric arrangement, the energy of the atom  and the energy of the field independently are equal at times before and after the scattering event. 

By contrast, an atom traversing the light crystal in the Doppler-shifted Laue geometry virtually absorbs a photon of frequency greater than that of the photon emitted via stimulation. The energy gained by the atom in the form of transverse kinetic energy is equal to the energy lost in the field by the unequal exchange of the field photons whose energies differ by $\hbar\delta$. From the spectroscopic point of view, the atom is involved in a virtual two-photon Raman scattering event as the virtually absorbed photon has a frequency different than that of the photon stimulating the emission \cite{RamanNath}. The deflected atom gains a transverse velocity proportional to the perpendicular momentum arising from the collision of the photon scattered $\pi$ radians by the recoil of the atom. The scattering is a two-photon virtual process since the field frequencies are far from resonance with the upper electronic state of the atom.  The atom's initial 0 transverse momentum is increased by $2\hbar k$, although the internal electronic energy of the atom is left unaltered. The direct product state is increased as the mechanical energy is increased by $2\hbar^2k^2/m$ while the atom is left in the electronic ground state. The direct product states are as follows:

\begin{eqnarray*}
\text{Undeflected(Incoming):} \quad\quad \ket{g.s.} &\otimes& \ket{p_\perp=0},   \\
 \text{Deflected(Diffracted):} \quad \quad \ket{g.s.} &\otimes& \ket{p_\perp=\sqrt{2m\hbar \delta}}.\\
\end{eqnarray*}

\section{Equations}

Off-resonant stimulated Raman scattering describes an exchange of energy from one field to the field counterpropagating with the incident atomic beam as the intermediary where adiabatic elimination of virtual transitions leads to a semiclassical formulation of the equations of motion. The fields, $\Omega$ and $\Omega + \delta$, are both far detuned by $\Delta$ from the
electronic ground to the excited state transition frequency $\omega$. The counterpropagating fields are detuned from each other such that their frequency difference, $\delta$, will later define a new set of direct product ground states for the atom, in which electronically the atom is in the ground state while mechanically it has transverse momentum $p_{\perp}=2n\hbar k$ for integer values of $n$.

In the Bragg regime there is only one pair of resonant states for each $\delta$. Far from the scattering fields, the atom's transverse momentum is either $0$ or $2\hslash k$ and off-resonant final state populations are negligible. Off-resonant final states are accessible in the Raman-Nath regime where the time the atom spends in the field is small with respect to the inverse of the recoil frequency of the Bragg diffracted atom; however, this regime is easily suppressed by a low atomic beam speed and a large laser waist.

\begin{widetext}
\subsection{Atom-optic equations of motion}

To derive atom-optic equations of motion, we start in the frame of the atom in which it encounters two counterpropagating pulses as in pump-probe spectroscopy. The two pulses are nearly identical in frequency ($\Omega$ and $\Omega+\delta$) and are far off-resonance with respect to the atom's electronic level separation ($\Delta=\omega-\Omega$). The two Rabi frequencies are denoted by $r_1(t)$ (for field at frequency $\Omega$) and $r_2(t)$ (for field at frequency $\Omega+\delta$). As we assume that the longitudinal speed of the atom is constant throughout the propagation within the laser-covered area, the time variation of $r_1(t)$ and $r_2(t)$ reflect the spatiotemporal variation of the counterpropagating beams. We write the Schr\"{o}dinger equation using the dipole operator to couple the electronic states to the motion in position representation where  $a_1(x,t)$ and $a_2(x,t)$ stand for the ground and excited electronic state amplitudes. We then transform the equations of motion into a momentum basis where the momentum amplitudes $a_{1}(p,t)$ and $a_{2}(p,t)$ are chosen to represent the deflected versus undeflected atoms. Following is a derivation of the equations of motion:

\begin{subequations}
\begin{eqnarray}
i\hbar\dot a_1(x,t)=-\frac{\hbar^2}{2m}\nabla^2 a_1(x,t)-\frac{1}{2}\hbar\omega a_1(x,t)
+\hbar[r_1(t) e^{-i\Omega t}e^{ik x}+r_2(t) e^{-i(\Omega+\delta)t}e^{-ik x}+c.c.]a_2(x,t),\label{adot1}\\
i\hbar\dot a_2(x,t)= -\frac{\hbar^2}{2m}\nabla^2 a_2(x,t)+\frac{1}{2}\hbar\omega a_2(x,t)
+\hbar[r_1(t) e^{-i\Omega t}e^{ik x}+r_2(t) e^{-i(\Omega+\delta) t}e^{-ik x}+c.c.]a_1(x,t). \label{adot2} 
\end{eqnarray}
\end{subequations}
\noindent 
Inserting the identity $\int_{-\infty}^{+\infty}\ket{p'}\bra{p'}dp'$ into Eq. (\ref{adot2}) and performing a Fourier transform yields the following:

\begin{equation}
\label{momentum2}
i\dot a_2(p,t)=\frac{p^2}{2\hbar m}a_2(p,t)+\frac{\omega}{2}a_2(p,t)+\int\int\braket{p}{x} [r_1(t) e^{-i\Omega t}e^{ik x}+r_2(t) e^{-i(\Omega+\delta) t}e^{-ik x}+c.c.]\braket{x}{p'} \braket{p'}{a_1(x,t)}dp'dx. 
\end{equation}

\noindent  Rewriting the last term of Eq. (\ref{momentum2}) by collecting exponents of equal spatial phase, we find 

\begin{equation}
\label{phaseterm}
\int\int\braket{p}{x}[r_1^*(t) e^{i\Omega t}e^{-ik x}+r_2(t) e^{-i(\Omega+\delta)t}e^{-ik x}+r_1(t) e^{-i\Omega t}e^{ik x}+r_2^*(t) e^{i(\Omega+\delta) t}e^{ik x}]a_1(p',t)dp'dx. 
\end{equation}

\noindent Introducing $R(t)$ and $R^*(t)$ as the positive and negative spatial phase terms, we obtain

\begin{subequations}
\label{phase}
\begin{eqnarray}
R(t)&=&r_1^*(t)e^{i\Omega t}+r_2(t) e^{-i(\Omega+\delta)t}, \\
R^*(t)&=&r_1(t) e^{-i\Omega t}+r_2^*(t) e^{i(\Omega+\delta)t}.
\end{eqnarray}
\end{subequations}

\noindent Substituting Eqs. (\ref{phase}) into Eq. (\ref{phaseterm}), we find

\begin{equation}
\label{lastterm}
\int\int\frac{e^{ix\cdot(p'-p)/\hbar}}{2\pi\hbar}(R(t) e^{-ik\cdot x}+R^*(t) e^{ik \cdot x})a_1(p',t)dp'dx =\hbar R(t)a_1(p+\hbar k,t)+\hbar R^*(t) a_1(p-\hbar k,t).
\end{equation}

\noindent Collecting the terms of Eq. (\ref{momentum2}) and dividing by $i\hbar$ we find the following equations of motion in the momentum basis:

\begin{subequations}
\label{momamp}
\begin{eqnarray}
\dot{a}_1(p,t)&=&i\left(-\frac{p^2}{2m\hbar}+\frac{\omega}{2}\right)a_1(p,t)
-iR(t) a_2(p+\hbar k,t)-iR^*(t)a_2(p-\hbar k,t), \\
\dot{a}_2(p,t)&=&i\left(-\frac{p^2}{2m\hbar}-\frac{\omega}{2}\right)a_2(p,t) -iR(t) a_1(p+\hbar k,t)-iR^*(t) a_1(p-\hbar k,t).
\end{eqnarray}
\end{subequations}

\noindent Introducing a momentum-interaction picture with the following redefined momentum amplitudes,

\begin{subequations}

\begin{eqnarray}
a_1(p,t)= \tilde a_1(p,t)e^{i(-\frac{p^2}{2m\hbar}+\frac{\omega}{2})t},\\
a_2(p,t)= \tilde a_2(p,t)e^{i(-\frac{p^2}{2m\hbar}-\frac{\omega}{2})t},
\end{eqnarray}
\end{subequations}
we find the following form:

\begin{subequations}
\begin{eqnarray}
\dot{\tilde a}_1(p,t)&=&
-iR(t)\tilde a_2(p+\hbar k,t)e^{i(-\frac{p k}{m}-\frac{\hbar k^2}{2m}-\omega)t}-iR^*(t)\tilde a_2(p-\hbar k,t)e^{i(\frac{p k}{m}-\frac{\hbar k^2}{2m}-\omega)t}, \label{atilde}\\
\dot{\tilde a}_2(p,t)&=&
-iR(t)\tilde a_1(p+\hbar k,t)e^{i(-\frac{p k}{m}-\frac{\hbar k^2}{2m}+\omega)t}-iR^*(t)\tilde a_1(p-\hbar k,t)e^{i(\frac{p k}{m}-\frac{\hbar k^2}{2m}+\omega)t}.
\end{eqnarray}
\end{subequations}

\noindent In order to eliminate $\tilde a_2(p,t)$ and find $\dot{\tilde a}_1(p,t)$ in terms of $\tilde a_1(p,t)$ alone, we integrate $\dot{\tilde a}_2(p,t)$, expand $R(t)$ and $R^*(t)$ with what they represent, and split the integrals:

\begin{equation}
\begin{aligned}    
\int\dot{\tilde a}_2(p,t)dt
=&-\int i r_1(t) e^{-i\Omega t}\,\tilde a_1(p-\hbar k,t)e^{i(\frac{p k}{m}-\frac{\hbar k^2}{2m}+\omega)t}dt
-\int ir_1^*(t) e^{i\Omega t}\,\tilde a_1(p+\hbar k,t)e^{i(-\frac{p k}{m}-\frac{\hbar k^2}{2m}+\omega)t}dt \\
&-\int
ir_2(t) e^{-i(\Omega+\delta)t}\,\tilde a_1(p+\hbar k,t)e^{i(-\frac{p k}{m}-\frac{\hbar k^2}{2m}+\omega)t}dt
-\int ir_2^*(t) e^{i(\Omega+\delta)t}\,\tilde a_1(p-\hbar k,t)e^{i(\frac{p k}{m}-\frac{\hbar k^2}{2m}+\omega)t}dt.
\end{aligned}
\end{equation}

\noindent After integration, and dropping terms with very large denominators namely $\omega+\Omega$, recalling that $\Delta=\omega-\Omega$, and assuming $\frac{p k}{m} << \Delta$ and $\frac{\hbar k^2}{2m} << \Delta$, we obtain

\begin{equation}
\tilde a_2(p,t)=-\frac{r_2(t)}{\Delta}e^{i(-\frac{p k}{m}-\frac{\hbar k^2}{2m}+\Delta-\delta)t}\tilde a_1(p+\hbar k,t)
-\frac{r_1(t)}{\Delta}e^{i(\frac{p k}{m}-\frac{\hbar k^2}{2m}+\Delta)t}\tilde a_1(p-\hbar k,t).
\end{equation}

\noindent The equation for $\tilde a_1(p,t)$ is in terms of $\tilde a_2(p+\hbar k,t)$ and $\tilde a_2(p-\hbar k,t)$, so we first substitute these momenta into the equation for $\tilde a_2(p,t)$:

\begin{subequations}
\label{virtual}
\begin{eqnarray}
\tilde a_2(p+\hbar k,t)&=&-\frac{r_2(t)}{\Delta}e^{i(-\frac{p k}{m}-\frac{3\hbar k^2}{2m}+\Delta-\delta)t}\tilde a_1(p+2\hbar k,t)
+\frac{r_1(t)}{\Delta}e^{i(\frac{p k}{m}+\frac{\hbar k^2}{2m}+\Delta)t}\tilde a_1(p,t),\\
\tilde a_2(p-\hbar k,t)&=&-\frac{r_2(t)}{\Delta}e^{i(-\frac{p k}{m}+\frac{\hbar k^2}{2m}+\Delta-\delta)t}\tilde a_1(p,t)
-\frac{r_1(t)}{\Delta}e^{i(\frac{p k}{m}-\frac{3\hbar k^2}{2m}+\Delta)t}\tilde a_1(p-2\hbar k,t).
\end{eqnarray}
\end{subequations}

 \noindent Substituting Eqs. (\ref{virtual}) into $\tilde a_1(p,t)$, Eq. (\ref{atilde}), and eliminating rapidly oscillating terms of the form $\Omega +\omega$, we find an equation for $\tilde a_1(p,t)$ where the virtual excited electronic state has been adiabatically eliminated from the equations of motion:

\begin{equation}
\label{atildedot}
\begin{aligned}    
\dot{\tilde a}_1(p,t)=
&-i\frac{r_1^*(t)r_1(t)+r_2^*(t)r_2(t)}{\Delta}\tilde a_1(p,t) \\
&-i\frac{r_1^*(t)r_2(t)}{\Delta}e^{i(-\frac{2p k}{m}-\frac{4\hbar k^2}{2m}-\delta)t}\tilde a_1(p+2\hbar k,t) 
-i\frac{r_2^*(t)r_1(t)}{\Delta}e^{i(\frac{2pk}{m}-\frac{4\hbar k^2}{2m}+\delta)t}\tilde a_1(p-2\hbar k,t).
\end{aligned}
\end{equation}

\noindent Label the level light shift

\begin{equation}
S(t)=\frac{r_1^*(t)r_1(t)+r_2^*(t)r_2(t)}{\Delta}
\end{equation}

\noindent and introduce

\begin{equation}
\beta f(t)=\frac{r_1^*(t)r_2(t)}{\Delta}=\frac{r_2^*(t)r_1(t)}{\Delta},
\end{equation}
\noindent where $f(t)$ is normalized to unity over the pulse duration, $\int_{-\infty}^{+\infty}f(t)dt=1$, and $\beta$ quantifies the field intensity divided by the detuning, $\Delta$. The two-photon recoil frequency is given by
\begin{equation}
4 \frac{\hbar k^2}{2m}=\frac{2\hbar k^2}{m}=\omega_k.
\end{equation}

\noindent The Bragg condition is achieved by the detuning, $\delta=\omega_k$, such that the recoil energy of the atom is equal to that of the loss of energy in the field and Eq. (\ref{atildedot}) can be recast as

\begin{equation}
\label{momentcoupled}
\dot{\tilde a}_1(p,t)=
-iS(t)\tilde a_1(p,t) 
-i\beta f(t)e^{-i\frac{2p k}{m}t}\tilde a_1(p+2\hbar k,t)
-i\beta f(t)e^{i(\frac{2pk}{m}-2\delta)t}\tilde a_1(p-2\hbar k,t).
\end{equation}

\noindent Equation (\ref{momentcoupled}) infers the coupling to an infinite number of states in the electronic ground state with varying momenta. We introduce an atomic beam where $p_\perp=0$.  The allowed outgoing states carry momenta indexed in multiples of $p=\pm2\hbar k$ and the equation of motion can be further simplified. The level light shift for first order is identical for all momentum states and can, therefore, be dropped from the coupled equations. Further, we may drop the electronic indices as the upper state is always virtual in the limit of large detuning (see discussion in \cite{PhysRevA.36.2495}). Last, we introduce subscripts to indicate the momenta of the various states:

\begin{equation}
i\dot{\tilde a}_{n \hbar k}(t)=
\beta f(t) e^{-in \delta t}\tilde a_{(n+2)\hbar k}(t) 
+ \beta f(t) e^{i(n-2)\delta t}\tilde a_{(n-2)\hbar k}(t).
\end{equation}

\noindent This equation clearly couples an infinite number of states; however, we begin by retaining only resonant states. The resonant states are those without phase factors and represent the states and couplings which arise from energy conservation between mechanical states of the atom and the field. The resonant equations where all higher-order Bragg processes are neglected are as follows with ${\tilde a}_n = {\tilde a}(n\hslash k,t)$:

\begin{subequations}
\begin{eqnarray}
\label{eqres}
i\dot {\tilde a}_{0}(t)=\beta f(t) {\tilde a}_2(t), \\
i\dot {\tilde a}_{2}(t)=\beta f(t) {\tilde a}_0(t). 
\end{eqnarray}
\end{subequations}

\noindent Including the next-order Bragg processes, the states ${\tilde a}_{-2}$ and ${\tilde a}_{4}$,

\begin{subequations}
\label{eqbare}
\begin{eqnarray}
i\dot {\tilde a}_{-2}(t)&=&\beta f(t) e^{i2\delta t}{\tilde a}_{0}(t), \\
i\dot {\tilde a}_{0}(t)&=&\beta f(t) e^{-i2\delta t}{\tilde a}_{-2}(t) +\beta f(t) {\tilde a}_2(t), \\
i\dot {\tilde a}_{2}(t)&=&\beta f(t) e^{-i2\delta t}{\tilde a}_{4}(t)+\beta f(t) {\tilde a}_0(t), \\
i\dot {\tilde a}_{4}(t)&=&\beta f(t) e^{i2\delta t}{\tilde a}_{2}(t),  
\end{eqnarray}
\end{subequations}

\noindent where the couplings to the ${\tilde a}_{-4}(t)$ and ${\tilde a}_{6}(t)$ terms, each of which in turn couple to higher-order states, are neglected symmetrically. 

Our last transformation of basis is performed to remove exponentials in order to speed the computational calculations. For the resonant states, we set 

\begin{subequations}
\label{dressed4}
\begin{eqnarray}
d_{-2}(t)={\tilde a}_{0}(t)e^{-i2\delta t},\\
d_0(t)={\tilde a}_{0}(t), \\
d_2(t)={\tilde a}_{2}(t), \\
d_{4}(t)={\tilde a}_{2}(t)e^{-i2\delta t},
\end{eqnarray}
\end{subequations}

\noindent which leads to a truncated matrix equation: 

\begin{equation}
i\dot d(t) =\left( 
\begin{array}{cccc}
2\delta & \beta f(t) & 0 & 0 \\ 
\beta f(t) & 0 & \beta f(t) & 0 \\ 
0 & \beta f(t) & 0 & \beta f(t)\\ 
0 & 0 & \beta f(t) & 2\delta
\end{array}
\right) d(t).
\end{equation}

\noindent For each pair of higher-order processes, the expressions for $d_n(t)$ include appropriate exponential terms such that the transformations lead to matrices containing only terms proportional to $\beta f(t)$ and $\delta$.

\subsection{Extended dressed basis and its consequence}

Most solutions to the equations of motion are found solving the four truncated equations of motion shown by Eq. (\ref{dressed4}). When the methods of approximation for the reduced four-level problem no longer suffice, we find that Eq. (\ref{extended}) gives arbitrarily precise solutions to the equations of motion in the form of a set of countably infinite coupled differential equations: 

\begin{equation}
\label{extended}
\newline
i
\left( 
\begin{array}{c}
\dots\\\dot{d}_{-8}(t)\\\dot{d}_{-6}(t)\\\dot{d}_{-4}(t)\\\dot{d}_{-2}(t)\\\dot{d}_{0}(t)\\\dot{d}_{2}(t)\\\dot{d}_{4}(t)\\\dot{d}_{6}(t)\\\dot{d}_{8}(t)\\\dot{d}_{10}(t)\\\dots\\
\end{array}
\right) =
\left( 
\begin{array}{rrrrrrrrrrrr}
\dots&\dots&\dots&\dots&\dots&\dots&\dots&\dots&\dots&\dots&\dots&\dots\\
\dots&20\delta&\beta f(t)&0&0&0&0&0&0&0&0&\dots\\
\dots&\beta f(t)&12\delta&\beta f(t)&0&0&0&0&0&0&0&\dots\\
\dots&0&\beta f(t)&6\delta&\beta f(t)&0&0&0&0&0&0&\dots\\
\dots&0&0&\beta f(t)&2\delta&\beta f(t)&0&0&0&0&0&\dots\\
\dots&0&0&0&\beta f(t)&0&\beta f(t)&0&0&0&0&\dots\\
\dots&0&0&0&0&\beta f(t)&0&\beta f(t)&0&0&0&\dots\\
\dots&0&0&0&0&0&\beta f(t)&2\delta&\beta f(t)&0&0&\dots\\
\dots&0&0&0&0&0&0&\beta f(t)&6\delta&\beta f(t)&0&\dots\\
\dots&0&0&0&0&0&0&0&\beta f(t)&12\delta&\beta f(t)&\dots\\
\dots&0&0&0&0&0&0&0&0&\beta f(t)&20\delta&\dots\\
\dots&\dots&\dots&\dots&\dots&\dots&\dots&\dots&\dots&\dots&\dots&\dots
\end{array}
\right) 
\left( 
\begin{array}{c}
\dots\\d_{-8}(t)\\d_{-6}(t)\\d_{-4}(t)\\d_{-2}(t)\\d_{0}(t)\\d_{2}(t)\\d_{4}(t)\\d_{6}(t)\\d_{8}(t)\\d_{10}(t)\\\dots\\
\end{array}
\right).
\end{equation}

Numerical solutions are easily produced provided the dimensionless detuning is not too small. Note that the on-diagonal coupling is found to be 
\begin{equation}
\label{betaorder}
\frac{n^2-2n}{4}\delta.
\end{equation}
\end{widetext}
\section{Numerical Solutions}

\subsection{Solutions to the bare four-state equations}

\begin{figure}
\includegraphics[width=8cm]{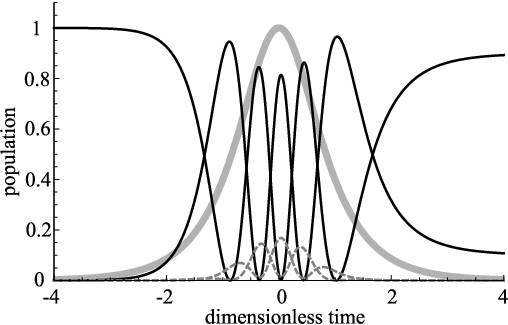}
\caption{Bare state probabilities with  $\delta=5$. The solid black lines represent the Bragg resonant  $|d_0(t)|^2$ and $|d_2(t)|^2$ states while the dashed gray lines represent the off-resonant states $|d_{-2}(t)|^2$ and $|d_{4}(t)|^2$. The thick gray line shows the curve of the pulse where maximum height is $\beta f(t)=\frac{1}{2}\sech(\frac{\pi t}{2})$. Dimensionless time is zeroed at the pulse maximum.}
\label{fig:sech4bare}
\end{figure}

The exact numerical solution of the four pairwise truncated bare state equations (\ref{eqbare}) is plotted in Fig. \ref {fig:sech4bare} for $f(t)=\frac{1}{2}\sech(\frac{\pi t}{2})$.  All values for time and frequency, $t$, $\delta$, and $\beta$, are dimensionless, where $t$ is in units of the pulse width and $\delta$ and $\beta$ are in units of the inverse pulse width. The initial states are unpopulated except for $d_0(-\infty)$, i.e., $p_{\perp}=0$ at $t=\infty$. Although the final states are resonant in the limit of large detuning, the off-resonant states are strongly populated during the pulse. It is clear that the off-resonant states are in fact contributing to the dynamics during the pulse. Neglecting to calculate off-resonant states during the interaction, $\dot {d}_{-2}(t)$ and  $\dot {d}_{4}(t)$ in Eqs. (\ref{dressed4}), clearly introduces error.

\begin{figure}
\includegraphics[width=8cm]{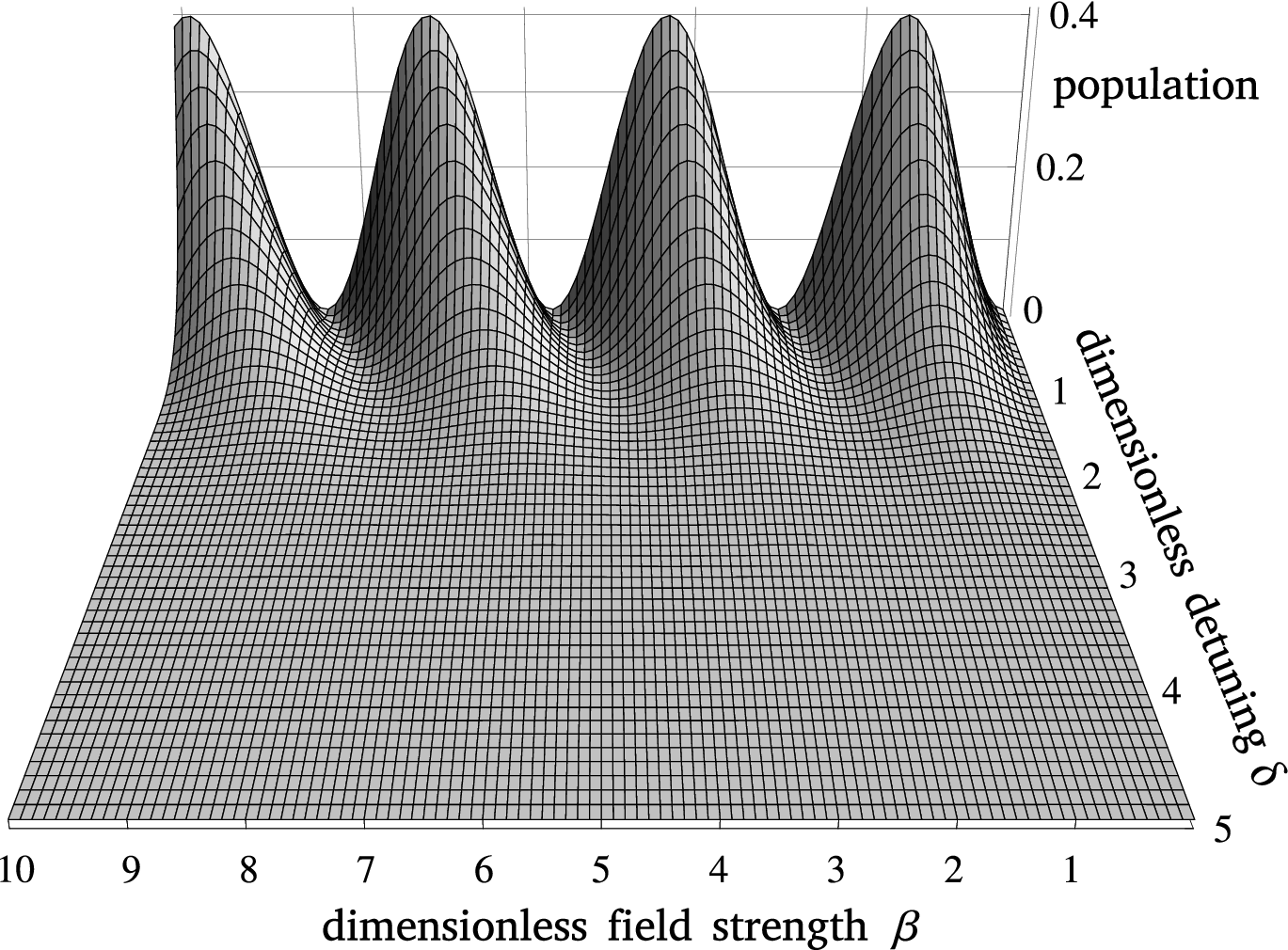}
\caption{The off-resonant state population, $|d_{4}(\infty)|^2$, is plotted against detuning and the dimensionless field strength, $\beta$. As $\delta$ grows larger, the final state population goes to zero. Note that we see evidence of Pendell\"osung.}
\label{fig:dress3d4}
\end{figure}

We seek the populations for $t=\infty$, long after the atom passes the light crystal when off-resonant states cease to play a role in the dynamics of the system. The two resonant states are designated by $|d_0(\infty)|^2$ and $|d_2(\infty)|^2$, where $\delta$'s and $\beta$'s are varied. Plotting the off-resonant probabilities, $|d_{4}(\infty)|^2$ with  $f(t)=\beta\frac{1}{2}\sech(\frac{\pi t}{2})$ against $\delta$ and $\beta$ we see evidence of the nonadiabaticity of higher-order states (see Fig. \ref{fig:dress3d4}). It is clear that the system exhibits Pendell\"osung as well, analogous to what is observed for altering the thickness of a crystal sample with Bragg scattered x rays. We shall demonstrate that the frequencies of these oscillations are strongly dependent on off-resonant states. However, if the detuning $\delta$ is too small, the Pendell\"osung is washed out as the outgoing states become indistinguishable as the corresponding spatial quantization tends toward a continuum of momentum states.

What must be determined is the number of states required in the calculation of the resonant population probability to assure adequate precision.

\section{The need for many states}

If the nonadiabatic contributions were  small enough, the various methods of approximation which only depend upon the resonant states and one pair of off-resonant states would suffice. We now demonstrate that such approximations break down drastically as the field strength is increased. For large enough detuning, the off-resonant populations can be made exponentially small; however, the effects of the higher-order momentum states are in fact necessary for correctly predicting the Pendell\"{o}sung frequencies of the resonant states and, therefore, the probabilities of finding atoms which are deflected as a function of field strength. The equations of motion are coupled, term by term, to higher off-resonant momentum states whose numbers are countably infinite. We seek the dependence of the number of states necessary for numerical calculation of outgoing Bragg resonant momentum states as a function of field strength for a fixed detuning.

As the field strength increases, the frequency of the Pendell\"{o}sung oscillations decreases. Although adequate detuning insures that the final population of off-resonant states is negligible, the effects of high-order off-resonant states during the pulse remain important. Using a hyperbolic secant pulse to calculate for the resonant two-state problem, we have the analytic Rosen-Zener \cite{rz} solution for the outgoing Bragg deflected beam,

\begin{equation}
d_{2_{RZ}}(\infty) = \sech (\delta) \sin(\beta),
\label{rzsolution}
\end{equation}

\noindent where $\delta$, the detuning, is usually kept fixed and $\beta$, the field amplitude, indicates the frequency of Rabi oscillations. When compared with the four-state solution for the Kapitza-Dirac system we are analyzing, this solution is inadequate. At low field strengths and large detuning, the frequency is identical; however, as the field strength is increased, the frequency drops precipitously. 

\begin{figure}
\includegraphics[width=8cm]{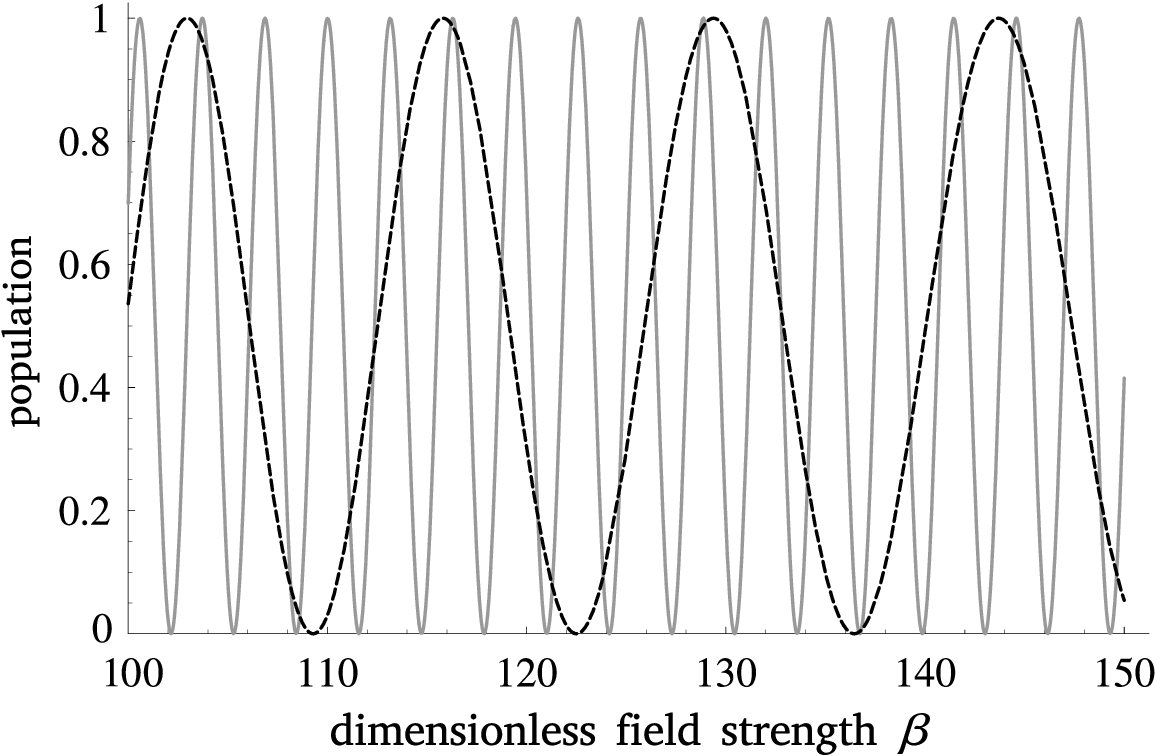}
\caption{Plots of the fraction of the deflected population versus the field strength $\beta$. The solid gray line is the Rosen-Zener solution, $|d_{2_{RZ}}(\infty)|^2$,  and the dashed black curve is the 14-state solution, $|d_{2_{(14 \text{ states})}}(\infty)|^2$. The deviation between the 14-state and higher solutions is less than 1\% and therefore represents a reasonably precise physical solution.}
\label{fig:142}
\end{figure}

In Fig. \ref{fig:142}, we have plotted the solutions to $|d_{2_{RZ}}(\infty)|^2$ and $|d_{2_{(14 \text{ states})}}(\infty)|^2$ to show the variation of phase and period. The dimensionless field strength ranges from 100 to 150, a region where both Pendell\"osung frequencies appear to be stable. Notice the Pendell\"osung frequency of the 14-state solution is roughly $1/4$ the frequency of the 2-state Rosen-Zener solution. For the same range of field intensity, the solutions calculated with fewer states are in fact monotonically greater in frequency until the 2-state solution is reached. For 16-state solutions and greater, for field intensities in the same range, the frequency is stable. The 14-state solution in Fig. \ref{fig:142} is close to converging on the physical solution of the resonant momentum states, i.e., the correct Pendell\"osung frequency and phase.

\begin{figure}
\includegraphics[width=8cm]{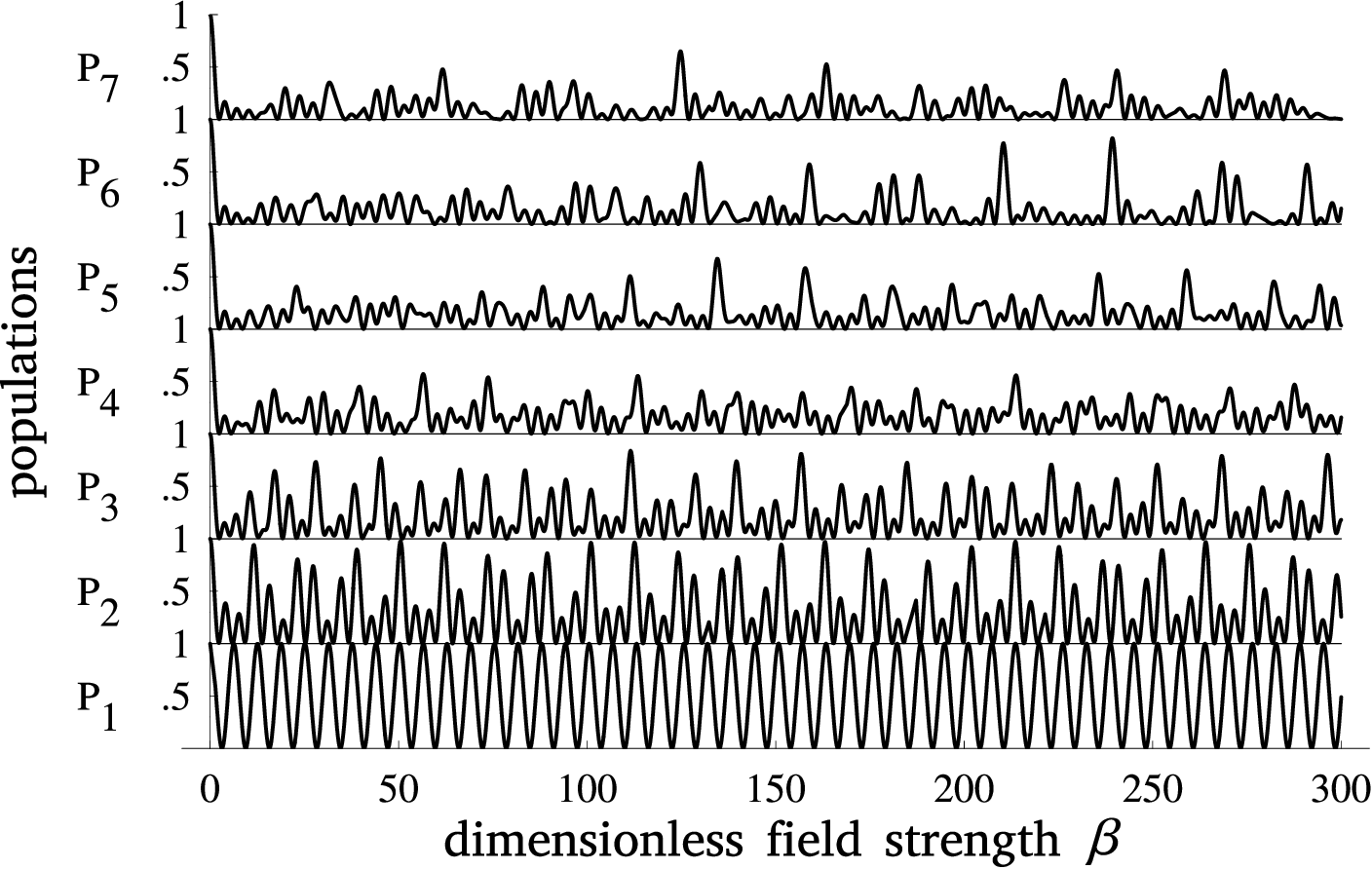}
\caption{The population fractions as a function of field strength $\beta$ for a detuning of $\delta=0.1$. Each curve is the solution to $|d_0(\infty)|^2$ for a successively higher number of states counted pairwise from 2.}
\label{fig:comparedpoint1}
\end{figure}

Although we do not vary the detuning in these plots, holding $\delta=5$, it should be noted that for a very small detuning, i.e., $\delta \ll 1$, we see a significant coupling to all high-order states. In fact, if the detuning is small, it may be impossible to predict the population of any state. Looking at Fig. \ref{fig:comparedpoint1} where the dimensionless detuning $\delta = 0.1$, the numerical solutions quickly become chaotic for field intensities of almost any magnitude. The final states may not be resonant, which indicates energy nonconservation reminiscent of the Raman-Nath regime \cite{chaos}. Physically, this chaotic regime is only speculative as stable computational solutions for $\delta \ll 1$ are practically impossible to find. In the Raman-Nath regime, where the interaction time is small compared to the inverse of the recoil frequency, uncertainty leads to many spatially quantized outgoing states. This can be understood by noting that when $\delta$ is small, the corresponding recoil momentum is small, and so the transverse momentum quantization is smeared and outgoing states tend toward an angular continuum where Bragg conditions are rendered moot. Therefore, to observe the atomic KDE, the detuning $\delta$ and the interaction time must both be large enough to allow for spatial quantization and energy conservation.

\begin{figure}
\includegraphics[width=8cm]{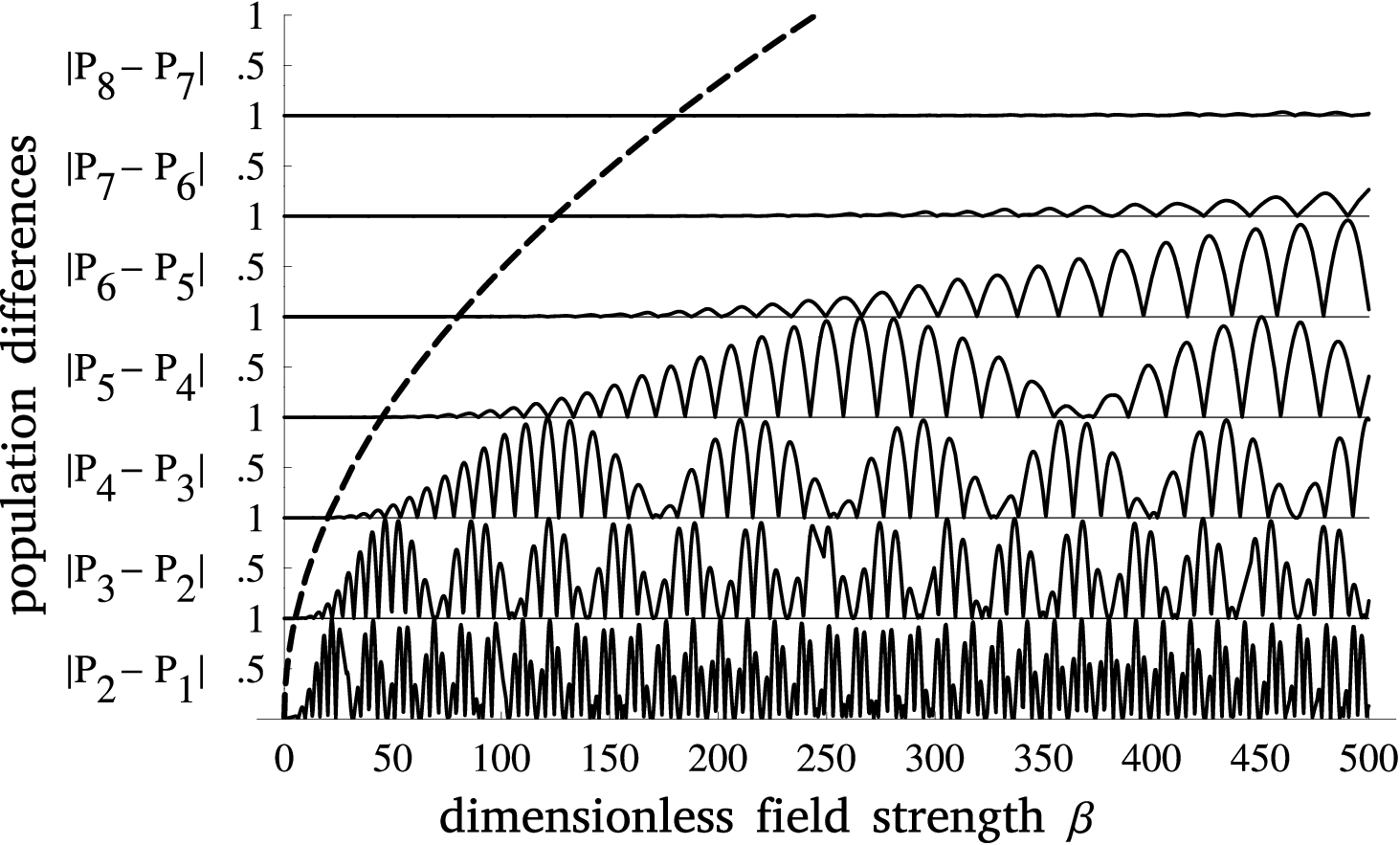}
\caption{Here the curve of $n = \sqrt{\beta /\delta}$ is shown as the dashed line superimposed on the graph of curves. Each of the seven curves represents the difference of population fraction for $|d_0(t=\infty)|^2$  when calculated respectively using $n$ and $n+2$ number of states, starting with the bottom-most curve on the graph with $n=2$.}
\label{comparison}
\end{figure}

Here, we posit an upper limit of the number of states, $n$, necessary to include in calculations achieving numerically stable solutions for a given $\beta$. The diagonal coupling between states given by Eq. (\ref{betaorder}) is suggestive of the following approximation for $n$,

\begin{equation}
\label{ordern}
n\propto\sqrt{\frac{\beta}{\delta}}, \delta > 1,
\end{equation}

\noindent where $n$ is the number of intermediate states used in calculating the final-state population probability of the deflected and undeflected atoms. 

To test Eq. (\ref{ordern}), in Fig. \ref{comparison} we plot Eq. (\ref{ordern}) over a set of graphs whose characteristics are similar to those of Fig. \ref{fig:comparedpoint1} in that each successively higher graph represents a higher order of Bragg scattering included in the calculation and the horizontal axis represents the dimensionless field amplitude. However, instead of graphing the population, the graphs measure $|P_{n+1}-P_n|$ using

\begin{equation}
\label{phase3d}
\begin{split}
P_{n}=|d_{(2n \text{ states})}(\infty)|^2,
\end{split}
\end{equation}

\noindent where $n$ is the order of the Bragg scattering included in the calculation involving $2n$ coupled equations. The bottom graph in Fig. \ref{comparison} consequently represents the difference between the two-state solution and the numerical four-state solution. It is clear from this plot that as the  dimensionless field amplitude increases, the two-state Rosen-Zener solution is inadequate. The region of Fig. \ref{comparison} above the curve of Eq. (\ref{ordern}) represents the physical solution where the variations from each higher and adjacent order are vanishingly small. 

With a detuning $\delta=5$, greater than 99.9\% of the atoms passing the light grating exit the field with a transverse momentum of $2\hbar$ or $0$. Although the resonant states are the only ones observed after passing the light crystal, the off-resonant states are populated during the scattering interaction with the light field analogous to multiple reflections of an x ray within a crystal before leaving a material sample. 

\begin{figure}
\includegraphics[width=8cm]{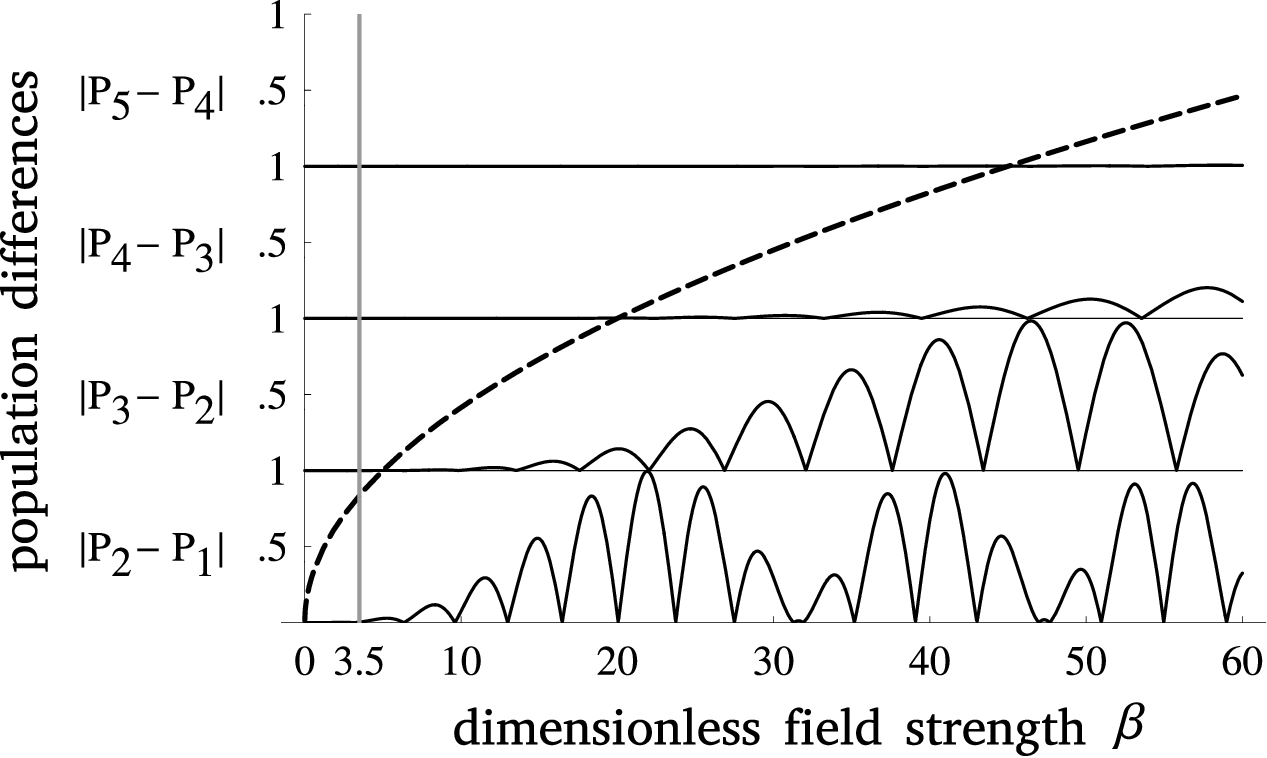}
\caption{To the left of the line $\beta = 3.5$, we see the population calculated by the two-level Rosen-Zener-like solution is sufficient and this region is therefore called the Rosen-Zener region. Notice that the curve denoting physical solutions misses this region as it is essentially asymptotic, so this region must be evaluated on its own.}
\label{regions}
\end{figure}

A precision calculation of the  Pendell\"osung leads us to ask how many states need be included in the calculation of the exiting states' populations, i.e., $|d_0(\infty)|^2$ and $|d_2(\infty)|^2$.  
Figure \ref{comparison} can be decomposed into three regions: the numerical, the unphysical, and the Rosen-Zener region as shown in Fig. \ref{regions}. The asymptotic solutions all fit in the unphysical region and may therefore only be of interest theoretically. The numerical and Rosen-Zener regions are both physical and therefore if one wishes to work with large dimensionless field amplitudes, one can always find the number of states required in the calculation from Eq. (\ref{extended}). 

\section{conclusions}

Figure \ref{comparison} clearly indicates that, unless the field strength is very low, the values for the final-state amplitudes predicted by the Rosen-Zener solutions will be incorrect. However, this can be useful in that one can design an experiment in which the dimensionless field amplitudes are small and the detuning is large enough that the quasianalytic two-state solutions are physical; that is, they properly predict the outgoing resonant state populations. If one needs to work in a regime of larger field amplitude, one can use Eq. (\ref{ordern}) to ensure that enough off-resonant states are included in the calculation to find physical solutions. Practically, it is more reasonable to utilize Eq. (\ref{extended}) starting with four states and increasing pairwise until the solution is stable. In contrast, if one wanted instead explicitly to probe the relevance of higher-order nonresonant states, one could design an experiment in the Bragg regime with $\beta$ of 1 MHz and $\delta$, the recoil frequency, of 100 kHz, as would be reasonable for a Bose-Einstein condensate. Here the phase of Pendell\"osung in the populated states would be noticeably different from the two-state Rosen-Zener solution and four-state approximations, well within the measurement capabilities of modern experiments.

The essential point of this paper is to make clear that the approximations presuming negligible contributions from the intermediate states are not reliable. The failure is not in the method of approximation, but in the underlying assumption that  one need consider only the pair of resonant states and the first pair of non-resonant states for precise determination of the phase and the frequency of the deflected atoms as a function of field strength. 

What we propose is for the physicist performing an experiment which utilizes spatial  quantization from the KDE to progressively include off-resonant intermediate states in the calculation of the final resonant state populations until the Pendell\"osung curve becomes stable. Not shown in this manuscript are calculations for a variety of pulse envelopes and a range of recoil frequencies, as the broad conclusions are the same; i.e., the final-state probabilities plotted against field strength become stable as more off-resonant intermediate states are added pairwise to the coupled differential equations numerically solved. Note as well, the Pendell\"{o}sung may seem sinusoidal at first glance, but careful inspection of the plots shows the period to vary with field strength. 

In the decades that have passed since the  pioneering experiments of atomic Kapitza-Dirac effect Bragg scattering, momenta transfers as high as $102\hbar k$ have been achieved by Kasevich, \textit{et al.} \cite{kasevich}. High-resolution interferometry based on Bragg diffraction is being carried out by groups such as Muller, \textit{et al.} \cite{muller} and Gerlich, \textit{et al.} \cite{nature} as well as work by Leeuwen, \textit{et al.} \cite{leeuwen}, where precise knowledge of the calculation of the resulting Pendell\"osung is critical to understanding system dynamics. 

\begin{acknowledgments}
I would like to thank P. R. Berman for his initial interest in this work, C. Genes, for his critique, and C. Aidala for her clarifications and generous support.

\end{acknowledgments}

\onecolumngrid

\end{document}